\documentclass[journal=jpclcd,
manuscript=letter,layout=twocolumn]{achemso} %to make two column
\usepackage[version=3]{mhchem}
\usepackage[T1]{fontenc}

\usepackage{array}
\usepackage{booktabs}
\usepackage{listings}
\usepackage{lmodern}
\usepackage{mathpazo}
\usepackage{microtype}
\usepackage{caption}
\usepackage{float}
\usepackage{setspace}
\usepackage{xkeyval}
\usepackage{color}
\usepackage{graphicx}

\usepackage{amsmath}
\usepackage{mathtools}
\usepackage{amssymb}
\usepackage{amsfonts}  
\usepackage{rotating}
\usepackage{multirow}
\usepackage{dcolumn}

\DeclareMathOperator\erfc{erfc}

\author{Joshua S. Kretchmer}
\email{jkretchm@caltech.edu}
\author{Garnet Kin-Lic Chan}
\email{gkc1000@gmail.com}
\affiliation[Caltech]{Division of Chemistry and Chemical Engineering, California Institute of Technology,
  Pasadena, California 91125}

\title{The fate of atomic spin in atomic scattering off surfaces}

\begin{document}

%%%%%%%%%%%%%ABSTRACT IMAGE%%%%%%%%%%%%%%
\begin{tocentry}
\includegraphics{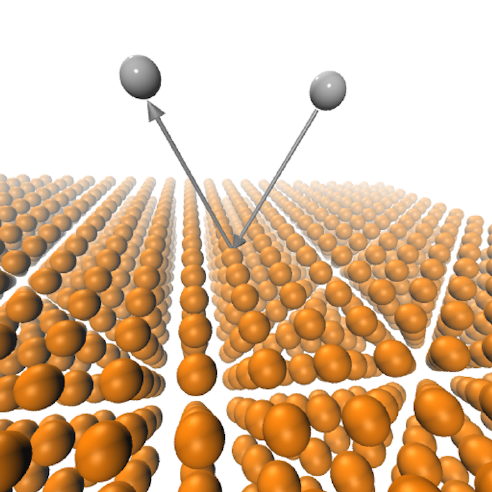}
\end{tocentry}
%%%%%%%%%%%%%%%%%%%%%%%%%%%%%%%%%

%%%%%%%%%%%%%ABSTRACT%%%%%%%%%%%%%%
\begin{abstract}
  We explore model electron dynamics of an atom scattering off a
  surface within the time-dependent complete active space self
  consistent field (TD-CASSCF) approximation.  We focus especially on
  the scattering of a hydrogen atom and its resulting spin-dynamics
  starting from an initially spin-polarized state.  Our results reveal
  competing electronic time-scales that are governed by the
  electronic structure of the surface as well as the character of the
  atom. The timescales and nonadiabaticity of the dynamics are
  reported on by the final spin-polarization of the scattered atom,
  which may be probed in future experiments.
\end{abstract}
%%%%%%%%%%%%%%%%%%%%%%%%%%%%%%%%%

Atomic scattering at surfaces is one of the fundamental events in surface chemistry.
However, in spite of the apparent simplicity, it involves many subtleties, including non-adiabatic effects,
and multiple mechanisms to dissipate translational energy into other modes.\cite{Kro08,Str98,Nie99,Ger01,Bun15}
Theoretical models have provided invaluable insights into how these processes take place.\cite{Bun15,Jan15,Nov15,Kro14,Miz08,Lin06,Bir04,Tim10,Lin06b,Miz05,Miz07,She06,Tra03,New69} For example, in the case of hydrogen atom (H-atom) scattering, 
calculations utilizing electronic friction approximations have pointed to the importance of electron-hole pair excitations as a dominant source of translational energy loss,\cite{Bun15,Jan15,Nov15,Kro14} while work based on time-dependent density functional theory or the time-dependent mean-field Newns-Anderson model have highlighted the importance of non-adiabaticity in chemicurrent experiments.\cite{Miz08,Lin06,Bir04,Tim10,Lin06b,Miz05,Miz07}

In this Letter,  we aim to understand the fate of the atomic spin during the surface scattering process.
We do so by simulating electron dynamics within a simplified model of the atom-surface dynamics.
As the hydrogen atom is the simplest  atom with an unpaired spin, it will be the main focus of our simulation parameters;
however, we will also tune our model to represent more complex spin polarized atoms.
Our simulations indicate that there are several classes of dynamical electron behaviour, which result in
different fates for the atomic spin. Measuring the final spin of the scattered atom in future spectroscopies will allow one to probe
the non-adiabatic character of the scattering process, as well as report on the electronic properties of the surface and the atomic-surface coupling.

%============IMAGE: FIG 1 SCHEMATIC=========
\begin{figure}[h!]
\includegraphics[scale=0.5]{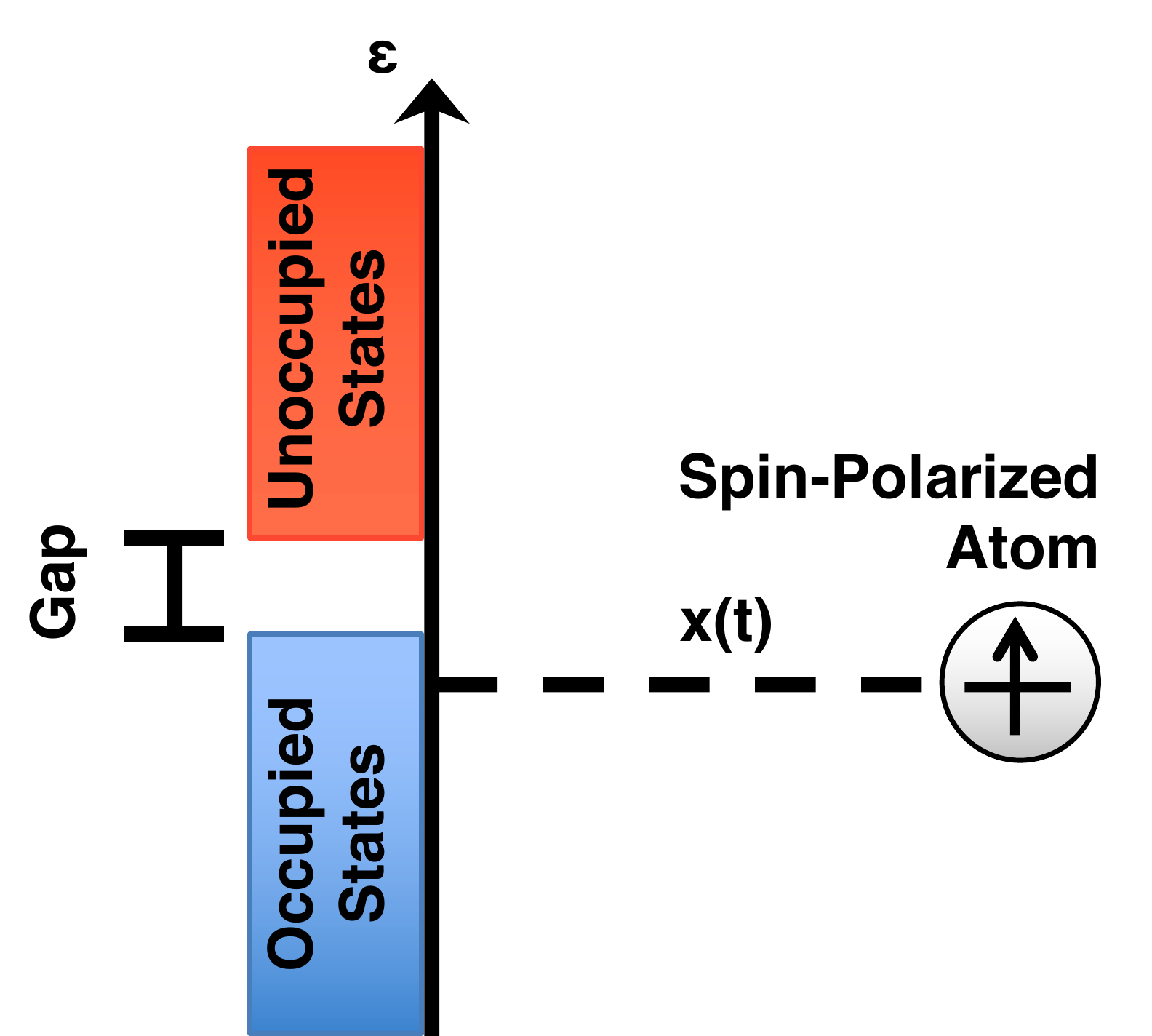}
\caption{\label{fig:schematic} Schematic illustration of the atom-scattering system. The system consists of an atom with a single, initially spin-polarized, electron a distance $x(t)$ away from the surface. The surface, which can be metallic or contain a gap, is initially in its electronic ground-state.}
\end{figure}
%========================================================

The model we will use is based on one previously parametrized to provide a realistic description of H-atom scattering, schematically illustrated in Fig. \ref{fig:schematic}.\cite{Miz08} The initially spin-polarized atom contains a single unpaired electron and is a distance $x(t)$ away from the surface; the surface is initially in its electronic ground-state. 
The electronic Hamiltonian for the model is
\begin{eqnarray}
\hat{H}(t)&=&\varepsilon_{\textrm{H}}(t)\hat{n}_{\textrm{H}}+U\hat{n}_{\textrm{H}\uparrow}\hat{n}_{\textrm{H}\downarrow}+\sum_{k}\varepsilon_k\hat{n}_k\nonumber\\
&+&\sum_{k\sigma}\left(V_{\textrm{H}k}(t)\hat{c}_{\textrm{H}\sigma}^\dag\hat{c}_{k\sigma}+h.c.\right),
\label{eqn:ham}
\end{eqnarray}
where $\varepsilon_{\textrm{H}}(t)$ and $U$ are the time-dependent single-particle energy and time-independent effective electron-electron repulsion on the atom, respectively. The term $\varepsilon_k$ is the single-particle energy for the $k$th state on the surface and $V_{\textrm{H}k}(t)$ is the time-dependent interaction between the $k$th state and the state on the atom, which is related to the width function, $\Gamma(t)$, through
\begin{equation}
\Gamma(t)=2\pi\sum_k|V_{\textrm{H}k}(t)|^2\delta\left(\varepsilon-\varepsilon_k\right).
\label{eqn:gam}
\end{equation}
The operator $\hat{c}_{i\sigma}^\dag$  ($\hat{c}_{i\sigma}^\dag$) creates (destroys) an electron of spin $\sigma$ in state $i$, where $i\in\{H,k\}$, $\hat{n}_{\textrm{H}\sigma}=\hat{c}_{\textrm{H}\sigma}^\dag\hat{c}_{\textrm{H}\sigma}$, and $\hat{n}_{\textrm{H}}=\hat{n}_{\textrm{H}\uparrow}+\hat{n}_{\textrm{H}\downarrow}$. The time-dependence of the parameters arises from the time-dependent motion of the atom, $x(t)$; the dynamical function $x(t)$ is obtained from pre-defined trajectories chosen to match experimental scattering conditions. The full details of the simulation model can be found in the Computational Methods section. By modifying the energy levels $\varepsilon_k$ one can simulate a range of surfaces including metallic, semiconducting, and insulating surfaces,
while changing $\varepsilon_H$ and $U$ allows for the treatment of different spin-polarized atoms.
 
We use the time-dependent complete active space self consistent field (TD-CASSCF) method to simulate the electron dynamics during the scattering process.\cite{Kat04,Cai05,Mir11,Sat13} The method has successfully been employed to simulate a variety of challenging electron dynamics problems including the multi-electron dynamics of molecules in intense laser fields,\cite{Kat04,Cai05,Mir11,Sat13,Hax12,Gre17,Lia17,Sat16} exciton dynamics in conjugated polymers,\cite{Mir11b,Fen17,Zha16,Zha13} and the non-equilibrium dynamics in molecular junctions.\cite{Lin15,Kre18}
It has several favourable characteristics in the current context, including the theoretical capability to reach a numerically exact answer
with increasing active space size, and the ability to accurately treat strongly correlated multi-electron processes. The code is implemented
using the \textsc{PySCF} package.\cite{Sun17}

%============IMAGE: FIG2 METAL SCATTERING=========
\begin{figure*}[h!]
\includegraphics{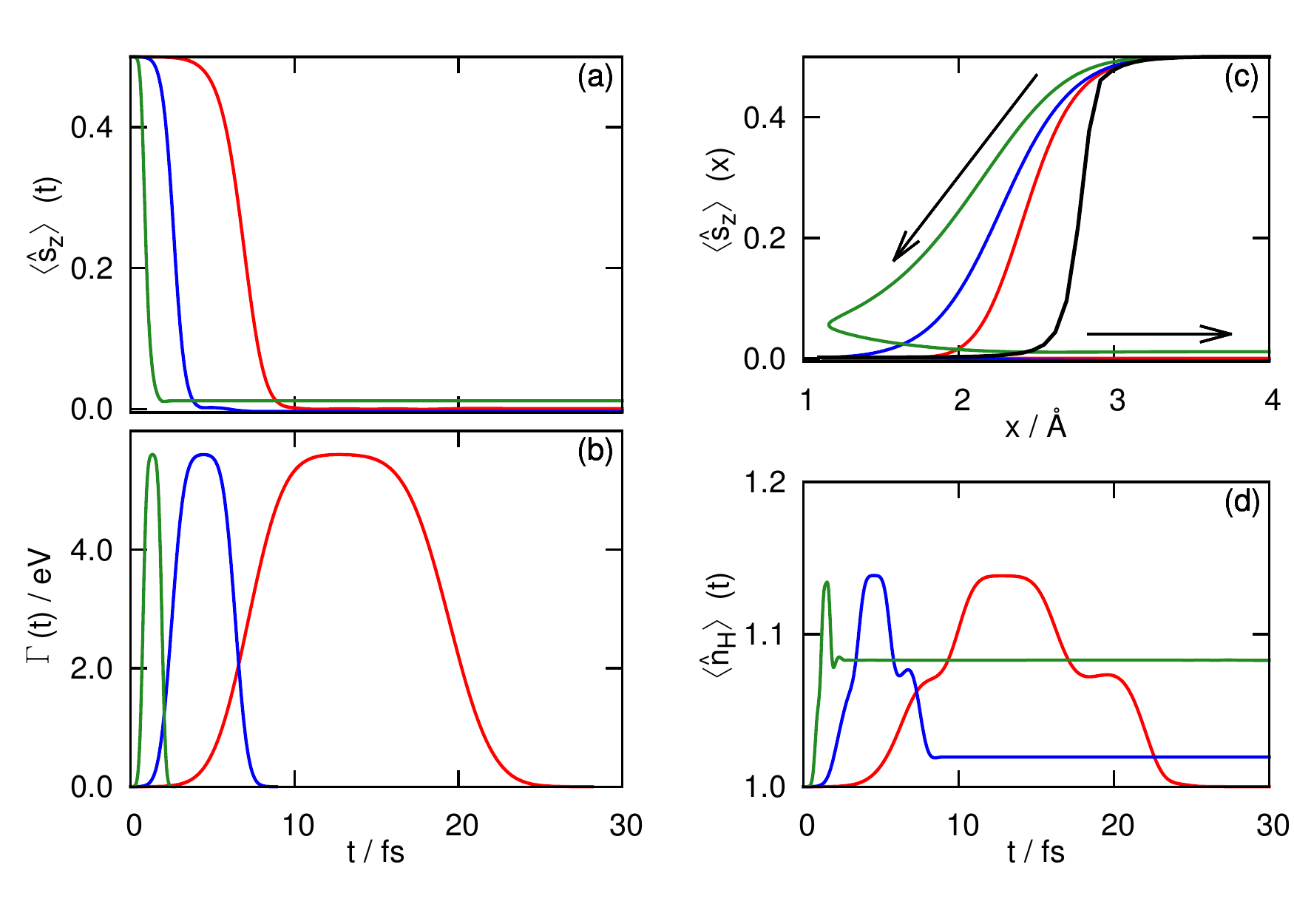}
\caption{\label{fig:metscat} Trajectories obtained from TD-CASSCF calculations of an initially spin-polarized H-atom scattering off of a copper surface for (a) the time-dependent expectation value of the spin-projection on the H-atom, $\langle\hat{s}_{\textrm{z}}\rangle(t)$, (b) the time-dependent width function, $\Gamma(t)$, (c) the position-dependent spin-projection on the H-atom $\langle\hat{s}_{\textrm{z}}\rangle(x)$, and (d) the time-dependent occupancy on the H-atom, $\langle\hat{n}_{\textrm{H}}\rangle(t)$. The colors indicate the scattering velocities with red corresponding to a slow trajectory, blue to a mid-level trajectory, and green corresponding to a fast trajectory, as described in the Computational Methods section. The black curve in panel (c) corresponds to the electronic ground-state value of the spin-projection and the arrows indicate whether the H-atom is moving towards or away from the surface.}
\end{figure*}
%========================================================

We start by exploring hydrogen atom scattering off of a copper surface.
We consider electron dynamics starting from different initial scattering conditions for the spin-polarized hydrogen atom.
Fig. \ref{fig:metscat}(a) presents the time-dependence of the expectation value of the spin-projection on the H-atom, $\langle \hat{s}_z\rangle(t)$, for a slow initial scattering velocity chosen to match previous experimental scattering conditions (red),\cite{Bun15} a mid-level initial scattering velocity (blue), and a fast initial scattering velocity (green); the full details of the scattering conditions can be found in the Computational Methods section. Fig \ref{fig:metscat}(b) shows the corresponding time-dependent width-function, $\Gamma(t)$, which is a measure of the interaction strength of the H-atom and the surface. As has been observed previously in the context of chemicurrent simulations,\cite{Miz08,Lin06} the spin on the H-atom depolarizes as it approaches the surface for all three scattering velocities. Additionally, our calculations show that the H-atom remains depolarized even at long-times after the H-atom
scatters far away from the surface. However, the fast trajectory does not fully depolarize during the scattering process; the time-scale for scattering is faster than the time-scale for the depolarization, such that the spin is only able to partially depolarize before the interaction between the H-atom and surface becomes too weak. This effect is highlighted in Fig. \ref{fig:metscat}(c), which displays the distance dependence of the expectation value of the spin-projection on the H-atom, $\langle \hat{s}_z\rangle(x)$. The black arrows indicate whether the H-atom is moving towards or away from the surface. In comparison to the mid-level and slow trajectories, the fast trajectory clearly does not fully depolarize prior to reaching the turning point for the H-atom (at 1.1 \AA).
Although the fast atom velocity is larger than that used in typical atomic scattering experiments, it is similar to those encountered in fast atomic diffraction and in ion scattering experiments.\cite{Xia12,Nie93,Bro07}

The incomplete spin depolarization arises from electronic non-adiabaticity, also illustrated in Fig. \ref{fig:metscat}(c). 
The black curve in Fig. \ref{fig:metscat}(c) shows the expectation value obtained from a ground-state CASSCF calculation, and thus corresponds to the electronically adiabatic result. 
The trajectories of all three scattering velocities show significant deviations from this ground-state curve indicating the presence of strong non-adiabatic effects induced by the inability of the electronic dynamics to respond sufficiently quickly to the timescale of the nuclear scattering process. The degree of non-adiabaticity, quantified by the deviation of the real-time trajectory from the electronically adiabatic result, increases with increasing scattering velocity.

To understand the mechanism behind the depolarization, Fig. \ref{fig:metscat}(d) presents the time-dependence of the expectation value of the occupancy on the H-atom, $\langle \hat{n}_{\textrm{H}}\rangle$(t). During the scattering process a small amount of charge density is transferred from the surface to the H-atom regardless of the scattering velocity.
However, the charge density is largely transferred back to the surface following the scattering process.
The degree of back transfer is determined by the scattering velocity, with the mid- and fast-level scattering velocities being sufficiently fast to out compete the time-scale for back transfer. However, the net transfer is very small and is not itself sufficient to depolarize the spin. 
Thus, we can conclude that the depolarization of the H-atom arises from charge fluctuations of the surface electrons, which can tunnel to and from the H-atom, rather than charge transfer itself. 

%============IMAGE: FIG3 GAPPED SCATTERING=========
\begin{figure}[h!]
\includegraphics{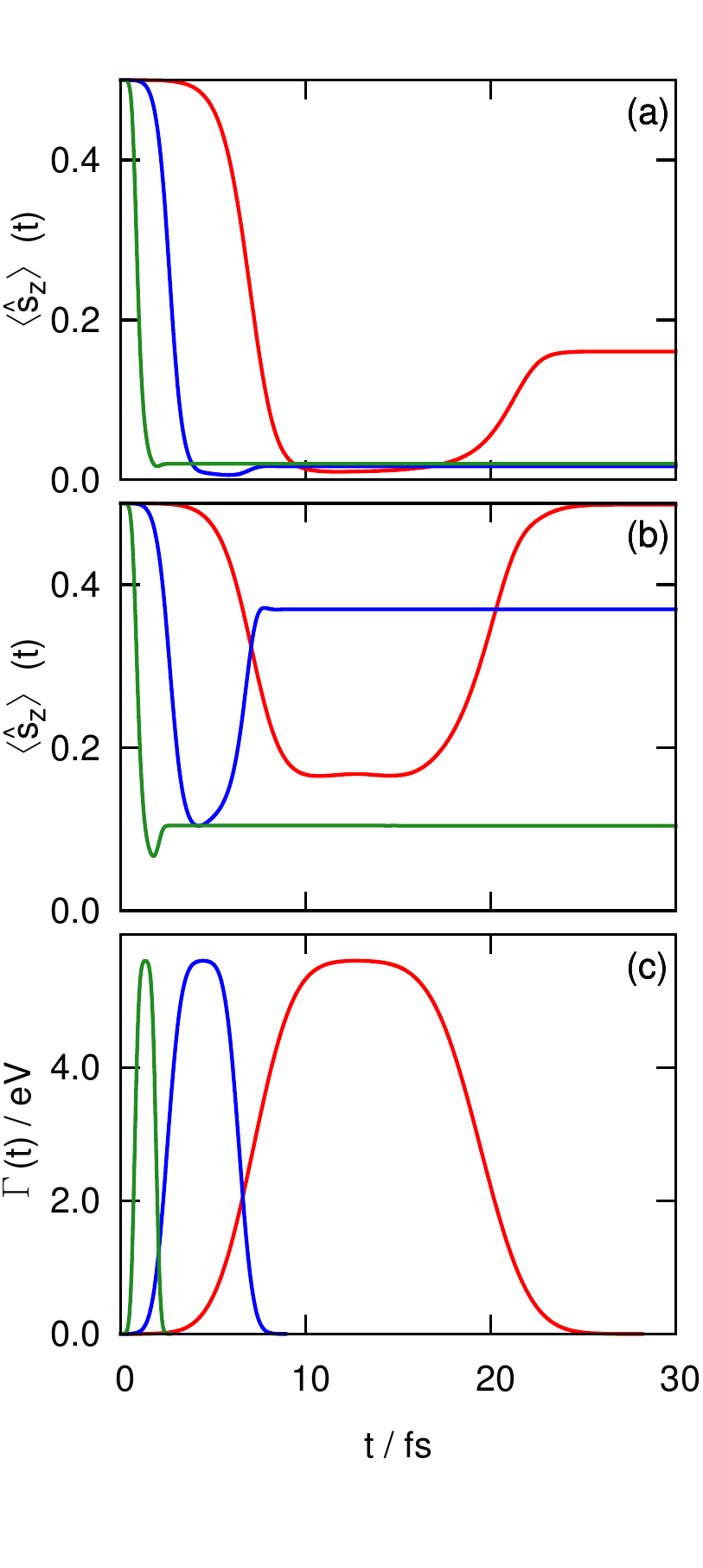}
\caption{\label{fig:gap} The time-dependent expectation value of the spin-projection, $\langle\hat{s}_{\textrm{z}}\rangle(t)$, on the H-atom for scattering at a surface with band-gap of size {\bf (a)} 0.3 eV and {\bf (b)} 1.0 eV. 
{\bf (c)} The corresponding time-dependent width function, $\Gamma(t)$, which is independent of the size of the band-gap. In all panels, the colors indicate the same scattering conditions used in Fig. \ref{fig:metscat}.
}
\end{figure}
%========================================================

The energy states in the metal surface introduce no time-scale of their own, as the density of states that we use is a flat function of energy
(wide-band limit).  To illustrate how modifying the surface
density of states  introduces additional time-scales, we next consider
an electronically gapped surface with gaps in the range of common semiconductors. Figure \ref{fig:gap} presents the spin trajectories for two such systems.
Figs. \ref{fig:gap}(a) and (b) plot the time-dependent spin-projection on the H-atom for a surface gap of 0.3 eV and 1.0 eV, respectively, and Fig. \ref{fig:gap}(c) plots the time-dependent width function, which is independent of the gap of the system; the colors in Fig. \ref{fig:gap} correspond to the same scattering conditions presented in Fig. \ref{fig:metscat}. In comparison to Fig. \ref{fig:metscat}(a), Fig. \ref{fig:gap} indicates that the introduction of an
electronic gap leads to a repolarization of the H-atom spin as the H-atom moves away from the surface following scattering. The degree of repolarization is controlled by both the size of the gap and the initial scattering velocity.

%============IMAGE: FIG4 GAPPED SCATTERING POSITION=========
\begin{figure}[h!]
\includegraphics{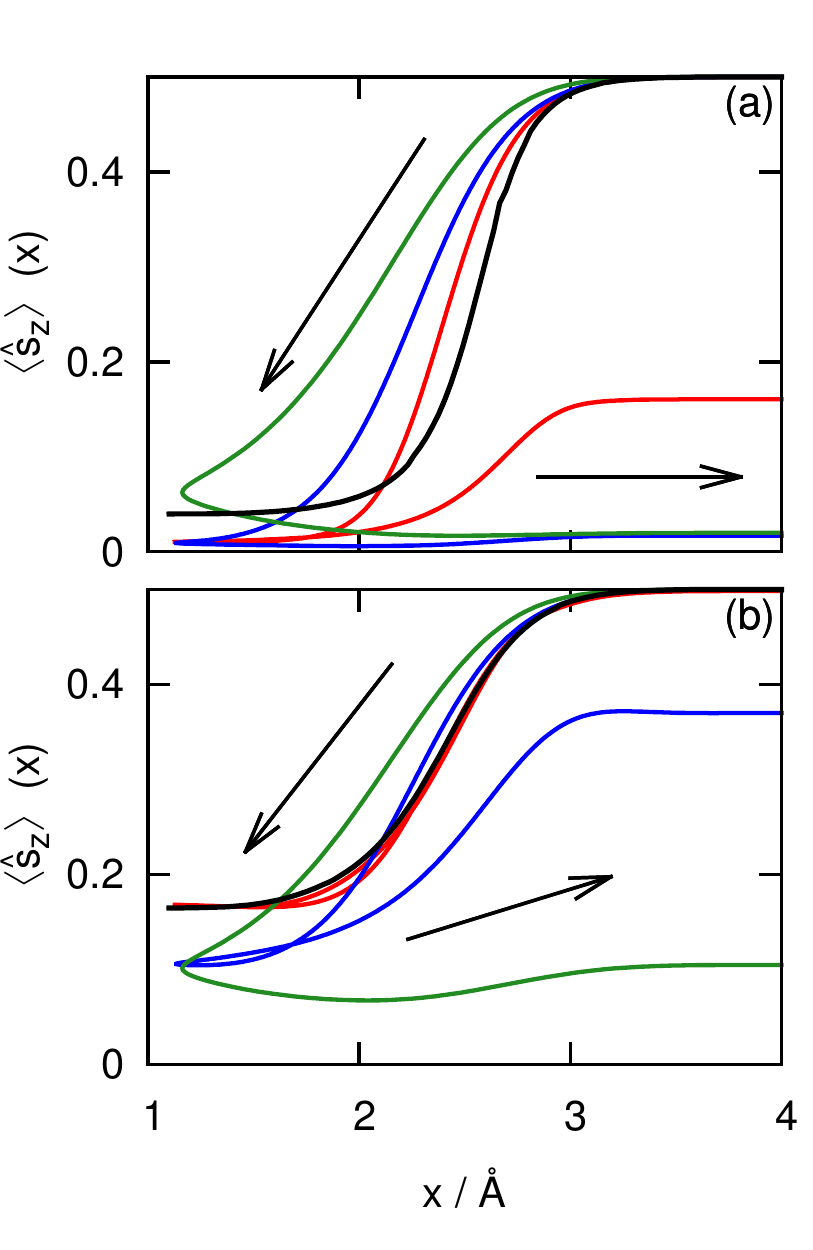}
\caption{\label{fig:gappos} The position-dependent expectation value of the spin-projection, $\langle\hat{s}_{\textrm{z}}\rangle(x)$, on the H-atom for scattering at a surface with band-gap of size {\bf (a)} 0.3 eV and {\bf (b)} 1.0 eV. In both panels, the colors indicate the same scattering conditions used in Fig. \ref{fig:metscat}.
}
\end{figure}
%========================================================

The trends observed in Fig. \ref{fig:gap} can be understood by examining Fig. \ref{fig:gappos}, which compares the position-dependent spin-projection obtained from the TD-CASSCF trajectories to the electronically adiabatic result obtained from a static CASSCF calculation given by the black curve; Figs. \ref{fig:gappos}(a) and (b) correspond to the surface containing a 0.3 eV and 1.0 eV gap, respectively, and the black arrows indicate whether the H-atom is moving towards or away from the surface. Non-adiabaticity arises from  time-dependent energy fluctuations surmounting the energy gap of the system.
Thus, the discrepancy between the dynamical trajectories and the adiabatic result decreases as the gap-size is increased and as the scattering velocity is decreased. The more adiabatic dynamics responds sufficiently quickly to the nuclear dynamics to then allow for the H-atom to repolarize
following scattering, which is most clearly seen for the slow trajectory (red) in Fig. \ref{fig:gappos}(b). The final degree of spin polarization on the H-atom increases as the ratio of the gap to the initial scattering velocity, for sufficiently large gaps. In the smaller gap system, however, even
the slowest velocity is not sufficiently slow to completely repolarize the H-atom.

%============IMAGE: FIGURE 5 Kondo=========
\begin{figure}[h!]
\includegraphics{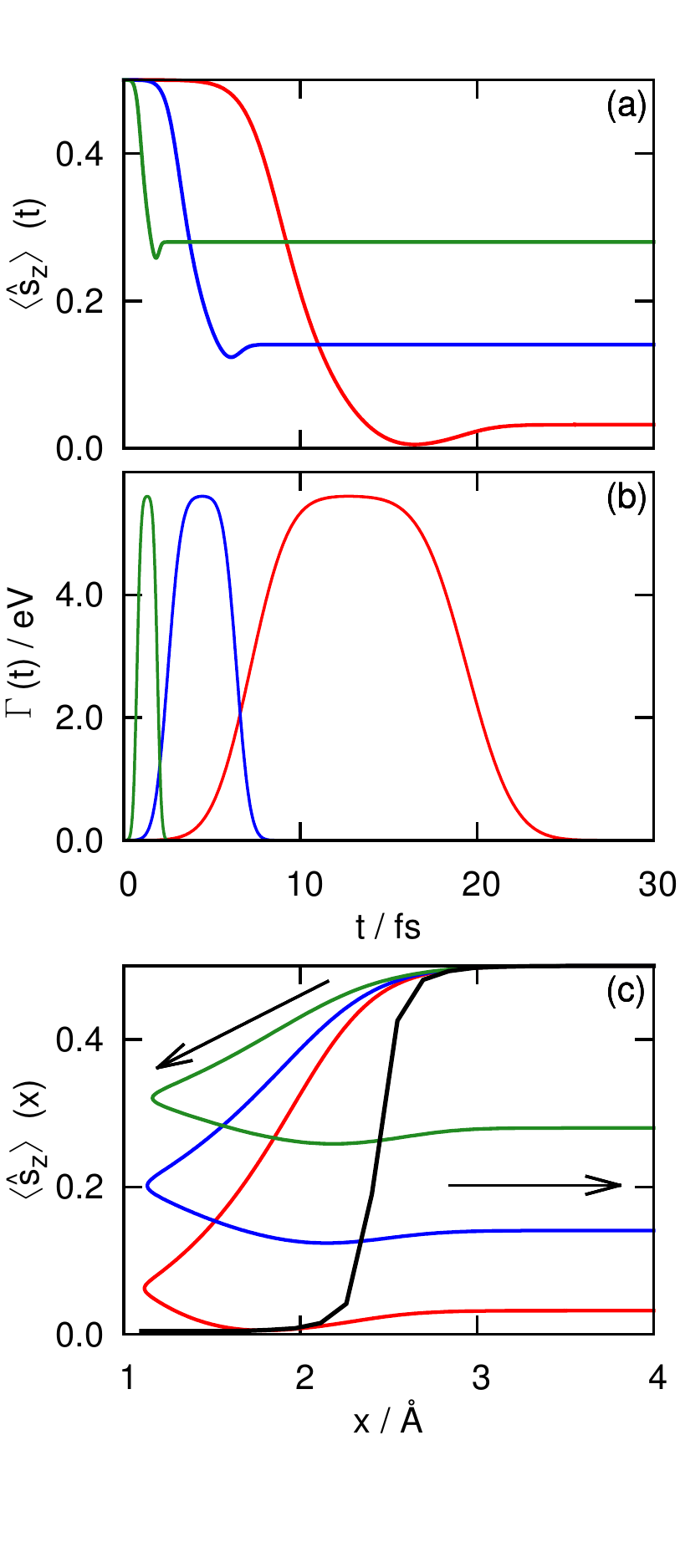}
\caption{\label{fig:kondo} Trajectories of an atomic species with a large value of the effective electron-electron repulsion, $U$, scattering off of a Cu surface for {\bf(a)} the time-dependent expectation value of the spin-projection, $\langle\hat{s}_{\textrm{z}}\rangle(t)$, {\bf(b)} the time-dependent width function, $\Gamma(t)$, and {\bf(c)} the position-dependent spin-projection on the H-atom $\langle\hat{s}_{\textrm{z}}\rangle(x)$. The colors indicate the same scattering conditions used in Fig. \ref{fig:metscat}.
}
\end{figure}
%========================================================

Finally, we consider the effects of varying an additional energy and time-scale in the scattering process that arises
when we consider atomic species other than hydrogen. Such atomic species will have
a different value of the electron-electron interaction parameter $U$. We consider an atom such as scandium, which, due to more localized atomic orbitals,
will have a higher value of $U$. A larger electron-electron interaction has been shown to introduce a slow timescale into the electron dynamics in model condensed-phase problems, associated with exchange.\cite{Nus15,Say16,Coh13,Coh11}
Fig. \ref{fig:kondo}(a) presents the spin-trajectories for an atomic species with a larger $U$ scattering off of a Cu surface for the same scattering conditions studied in Fig. \ref{fig:metscat}; Fig. \ref{fig:kondo}(b) plots the corresponding time-dependent width-functions. The initial time-scale for depolarization of the spin is clearly seen to increase in this system compared to H-atom scattering; the slow trajectory (red) in Fig. \ref{fig:kondo}(a) does not fully depolarize until $\sim$16 fs compared to the $\sim$10 fs for H in Fig. \ref{fig:metscat}(a). In fact, the time-scale for depolarization now becomes sufficiently slow in comparison to the time-scale of the scattering process for both the mid-level (blue) and fast (green) trajectories, that the atomic species no longer fully depolarizes. This means that one can preserve the initial atomic spin state even in metallic scattering if the atomic species has sufficiently localized atomic states.

The slower time-scale for depolarization is indicative of a higher degree of non-adiabaticity during surface scattering with the increased $U$, as indicated by Fig. \ref{fig:kondo}(c). The dynamical trajectories in Fig. \ref{fig:kondo}(c) exhibit large discrepancies from the electronically adiabatic result obtained from a static CASSCF calculation (black curve); the black arrows indicate whether the atom is moving towards are away from the surface. As before, the degree of non-adiabaticity increases with increasing scattering velocity, which is exemplified by the decrease in the extent of depolarization for faster scattering velocities.

In summary, our time-dependent complete active space self consistent field (TD-CASSCF)  simulations
of parametrized models of spin-polarized atomic scattering uncover 
 a variety of electron dynamics and timescales,  leading
to widely varying fates of the atomic spin. 
We find that  the degree to which the spin depolarizes
 sensitively reports on many aspects of the atom and surface electronic structure, such as the bandgap
and electron-electron interaction, as well as the timescales and non-adiabaticity of the electronic processes. Experiments to probe
non-adiabatic effects have in the past relied on energy loss by the scattering atom, where one has to disentangle
the contributions of nuclear mode coupling from that of electron-nuclear coupling. Intriguingly, our simulations suggest that measurements of the
final spin polarization may well provide a more direct spectroscopic route in future experiments to uncover the detailed electronic dynamics of atomic scattering off surfaces.

\section{Computational Methods}

The time-dependence of the terms for the electronic Hamiltonian of the scattering system defined in Eq. \ref{eqn:ham} are obtained from a previously utilized model for H-atom scattering such that,\cite{Miz08}
\begin{equation}
\varepsilon_{\textrm{H}}(t)=a_1+a_2\erfc\left[a_3\left(x(t)-a_4\right)\right]
\end{equation}
and
\begin{equation}
\Gamma(t)=b_1+b_2\erfc\left[b_3\left(x(t)-b_4\right)\right],
\end{equation}
where $x(t)$ is the time-dependent distance of the H-atom from the surface, Fig. \ref{fig:schematic}. The width function is defined in the wide-band limit and is independent of the energy of the surface states.\cite{Miz08}

\begin{table}
  \caption{Electronic parameters}
  \label{tab:params}
  \begin{tabular}{ll}
    \hline
    Parameter & Value \\
    \hline
    $a_1$ (eV) & -2.872 \\
    $a_2$ (eV) & -0.263 \\
    $a_3$ (\AA$^{-1}$) & 3.717  \\
    $a_4$ (\AA) & 1.729 \\
    $b_1$ (eV) & -8.020$\times10^{-5}$ \\
    $b_2$ (eV) & 2.805 eV \\
    $b_3$ (\AA$^{-1}$) & 1.796 \\
    $b_4$ (\AA) & 2.352 \\
    \hline
  \end{tabular}
\end{table}

The value of the parameters used in the manuscript are chosen to match the interaction between a H-atom and a Cu surface and are presented in Table \ref{tab:params}.\cite{Miz08} The effective electron-electron interaction, $U$ in Eq. \ref{eqn:ham}, is given by $U=4.827$ eV for H-atom scattering and $U=20.0$ eV for the scattering system involving an atomic species with a higher electron-electron interaction studied in Fig. \ref{fig:kondo}. In addition, the term $a_1=-9.475$ eV in Fig. \ref{fig:kondo} to place the system closer to the Kondo regime in which the slower electronic time-scale is most prevalent;\cite{Nus15,Say16,Coh13,Coh11} this value physically corresponds to introducing a bias to the Cu surface to change the Fermi-level relative to the single-particle state on the atomic species.

In this work we use a constant density of states for the surface such that the values of $\varepsilon_k$ in Eq. \ref{eqn:ham} are evenly spaced in the interval $[-10, 10]$ eV for the Cu surface and in the intervals $[-10, -E_{\textrm{gap}}/2.0]$ eV and $[E_{\textrm{gap}}/2,10]$ eV for the gapped surfaces studied in Figs. \ref{fig:gap} and \ref{fig:gappos}, where $E_{\textrm{gap}}$ is the size of the band-gap. An even density of states yields the following discretization of the width function,\cite{Wan13}
\begin{equation}
|V_{\textrm{H}k}(t)|^2=\frac{\Gamma(t)\Delta\varepsilon}{2\pi},
\end{equation}
where $\Delta\varepsilon$ is the energy spacing of the surface states.

\begin{table}
  \caption{Nuclear parameters}
  \label{tab:Hparams}
  \begin{tabular}{lccc}
    \hline
    Parameter & Slow & Mid-Level & Fast \\
    \hline
    $v_i$ (\AA\ / fs) & -0.23 & -0.66 & -2.19\\
    $v_f$ (\AA\ / fs) & 0.19  & 0.66 & 2.19\\
    $\alpha$ / $10^{-3}$ (fs$^{-1}$)& 4.13 & 8.27 & 20.67 \\
    $t_0$ (fs) & 12.70 & 4.45 & 1.33 \\
    \hline
  \end{tabular}
\end{table}

The dynamics of $x(t)$ are obtained from a simplified pre-defined trajectory, in which the velocity of the H-atom is given by
\begin{equation}
v(t)=\frac{v_f-v_i}{2}\tanh\left[\alpha\left(t-t_0\right)\right]+\frac{v_f+v_i}{2},
\label{eqn:Hvel}
\end{equation}
where $v_i$ and $v_f$ define the initial and final scattering velocities. The velocity can be integrated to yield
\begin{eqnarray}
x(t)&=&x_i-\frac{v_f-v_i}{2\alpha}\ln\left[\cosh\left(\alpha t_0\right)\right]\nonumber\\
&+&\left(\frac{v_f+v_i}{2}\right)t\nonumber\\
&+&\frac{v_f-v_i}{2\alpha}\ln\left[\cosh\left(\alpha(t-t_0)\right)\right].
\label{eqn:Hpos}
\end{eqnarray}
In this work, we present results from three sets of values for the parameters in Eqs. \eqref{eqn:Hvel} and \eqref{eqn:Hpos} corresponding to a slow, a mid-level, and a fast initial scattering velocity.
The values for the slowest initial scattering velocity are chosen to match previously measured experimental scattering velocities.\cite{Bun15} All remaining parameters are chosen such that the point of closest approach of the H-atom to the surface is 1.1 \AA\ given an initial position of $x_i = 4$ \AA\ from the surface at time zero.
The values of the parameters are presented in Table \ref{tab:Hparams}.
We note that the use of a pre-defined nuclear trajectory is a severe approximation to the true dynamics of the system. However, the use of pre-defined trajectories has proved successful for the simulation of chemisorption and we thus expect it to provide a reasonable description for the nuclear dynamics for the scattering systems investigated in this work.\cite{Miz08}

All TD-CASSCF calculations were found to converge with 128 surface states and an active space size of 7. The active space size is defined to include the single-particle state on the H-atom, and the HOMO through HOMO-2 and LUMO through LUMO+2 states of the surface. The equations of motion are evaluated using the fourth-order Runge-Kutta method with a time-step of 0.024 as. A regularization parameter of size $2.72\times10^{-7}$ eV is introduced, as has been done in previous applications of TD-CASSCF.\cite{Kre18}

%%%%%%%%%%%%%ACKNOWLEDGMENTS%%%%%%%%%%%%%%
\begin{acknowledgement}
  This work was supported by the US Department of Energy, Office of Science through DOESC0018140, and by the Simons Foundation via
  the Many Electron Collaboration. GKC is a Simons Investigator in Physics.
\end{acknowledgement}
%%%%%%%%%%%%%%%%%%%%%%%%%%%%%%%%%

%%%%%%%%%%%%%BIBLIOGRAPHY%%%%%%%%%%%%%%
\bibliography{hscat_biblio}
%%%%%%%%%%%%%%%%%%%%%%%%%%%%%%%%%

\end{document}